\documentclass[conference]{IEEEtran}
\IEEEoverridecommandlockouts
\usepackage{cite}
\usepackage{amsmath,amssymb,amsfonts}
\usepackage{graphicx}
\usepackage{textcomp}
\usepackage{xcolor}
\usepackage{subcaption}
\usepackage{algorithm}
\usepackage{algpseudocode}
\usepackage{hyperref} 
\usepackage{xcolor}
\definecolor{codegreen}{RGB}{0,128,0}
\definecolor{linkblue}{RGB}{0,90,156}

\def\BibTeX{{\rm B\kern-.05em{\sc i\kern-.025em b}\kern-.08em
    T\kern-.1667em\lower.7ex\hbox{E}\kern-.125emX}}
\begin{document}

\title{Distributed Quantum Error Mitigation: Global and Local ZNE encodings
}

\author{\IEEEauthorblockN{Maria Gragera Garces}
\IEEEauthorblockA{\textit{Quantum Software Lab, University of Edinburgh} \\
Scotland, UK \\
0009-0000-9018-7435}
}

\maketitle

\begin{center}
{\footnotesize\textit{Accepted at IEEE INFOCOM 2026 (QUNAP)}}
\end{center}

\begin{abstract}
Errors are the primary bottleneck preventing practical quantum computing.
This challenge is exacerbated in the distributed quantum computing regime, where quantum networks introduce additional communication-induced noise.
While error mitigation techniques such as Zero Noise Extrapolation (ZNE) have proven effective for standalone quantum processors, their behavior in distributed architectures is not yet well understood.
We investigate ZNE in this setting by comparing Global optimization (ZNE is applied prior to circuit partitioning), against Local optimization (ZNE is applied independently to each sub-circuit).
Partitioning is performed on a monolithic circuit, which is then transformed into a distributed implementation by inserting noisy teleportation-based communication primitives between sub-circuits.
We evaluate both approaches across varying numbers of quantum processing units (QPUs) and under heterogeneous local and network noise conditions.
Our results demonstrate that Global ZNE exhibits superior scalability, achieving error reductions of up to $48\%$ across six QPUs.
Moreover, we observe counterintuitive noise behavior, where increasing the number of QPUs improves mitigation effectiveness despite higher communication overhead.
These findings highlight fundamental trade-offs in distributed quantum error mitigation and raise new questions regarding the interplay between circuit structure, partitioning strategies, and network noise.
\end{abstract}


\section{Introduction}

Quantum computing promises efficient solutions to problems that are intractable for classical systems, yet practical algorithms require thousands to millions of qubits \cite{gidney2021factor}, far exceeding the qubit counts, connectivity, and noise limits of individual devices.
This gap has driven the emergence of distributed quantum computing (DQC), 
where circuits are partitioned across devices interconnected by quantum networks, 
mirroring classical computing's evolution toward distributed architectures and already being necessitated 
at small scales in systems like IBM's Heron chips and Xanadu's Aurora device.

This distribution of quantum compute over a quantum network will come with many challenges 
including new resouces to manage (entanglement) and new sources of noise \cite{campbell2024quantum}.

In this work, we explore how error mitigation, 
particularly Zero Noise Extrapolation (ZNE), a cheaper near-term alternative to quantum error correction, 
performs in a DQC setting. 
We investigate whether Global optimisation (applying ZNE before circuit partitioning) 
or Local optimisation (applying ZNE independently to each sub-circuit) represents a more effective approach. 

While error mitigation techniques have been extensively studied for standalone quantum processors, 
very little prior work exists at the intersection of error mitigation and DQC. 
Existing work relies on basic strategies such as routing computations to higher-quality qubits~\cite{chen2024noise}, rather than adapting sophisticated mitigation techniques to the distributed setting.
This work represents the first empirical exploration of how state-of-the-art error mitigation strategies interact with circuit partitioning and quantum network communication. 

Our experimental evaluation comprises over 3,500 simulations using real circuits from the MQT Bench suite~\cite{quetschlich2023mqt}, 
executed on Qiskit Aer, with custom noise models that capture both intra-device and inter-device errors. 

\section{Background}
\subsection{Distributed Quantum Circuits}
Current quantum computers face significant limitations in qubit count and connectivity, 
preventing the execution of many practical algorithms on a single device. 
DQC addresses this challenge by partitioning circuits across multiple quantum processors connected via a quantum network.

To distribute a quantum circuit, 
we first abstract it as a graph \footnote{An alternative representation is a hypergraph, which has been extensively studied in the literature \cite{barral2025review}. Hypergraphs naturally encode multi-way interactions and can directly represent multi-qubit gates such as the Toffoli. For simplicity, we adopt a graph abstraction in this work and decompose Toffoli gates accordingly.} 
where nodes represent qubits and edges represent two-qubit gates. 
Graph partitioners then partition this structure into sub-circuits, 
each executable on a separate quantum processor. 
Gates spanning partition boundaries must be replaced with non-local communication primitives such as teleportation (TP), 
which transfers quantum states using pre-shared entanglement over a quantum channel.
While these non-local gates enable inter-QPU quantum operations, 
they introduce additional computational overhead and communication-induced noise, 
expected to exceed intra-device noise~\cite{campbell2024quantum}.

\subsection{Zero Noise Extrapolation}
Zero Noise Extrapolation (ZNE) is a quantum error mitigation technique that estimates 
the noise-free expectation value of an observable without requiring additional qubits or quantum error correction~\cite{li2017efficient,temme2016error}. 
ZNE operates in two phases: 
\begin{enumerate}
\item Noise scaling: the circuit is executed at multiple artificially increased noise levels by gate folding (repeating gates to artificially increase depth) or pulse stretching (slowing gate pulses to amplify decoherence).
\item Extrapolation: the measured expectation values are fitted to a functional form (linear, polynomial, or exponential) and extrapolated to the zero-noise limit.
\end{enumerate}

In this work, we use Mitiq~\cite{larose2022mitiq} (an open-source Python toolkit for quantum error mitigation) to implement ZNE via unitary folding with linear extrapolation. 
The critical question we investigate is \textit{when} to apply ZNE in a distributed setting: before partitioning (Global optimisation) or after partitioning to individual sub-circuits (Local optimisation). 
This timing decision significantly impacts both the circuit structure preserved during mitigation and the overhead introduced by communication primitives.

\section{Methodology}

\subsection{Experimental Setup}

We evaluated our approaches using three quantum algorithms from the MQT Bench suite~\cite{quetschlich2023mqt}: 
Greenberger-Horne-Zeilinger (GHZ) state preparation, Deutsch-Jozsa (DJ), and W state preparation. 
These algorithms represent a diverse set of computational patterns with varying gate connectivity and circuit depth, 
enabling the assessment on various circuit structures.

All experiments were conducted using the Qiskit Aer simulator~\cite{qiskit2024} 
with custom noise models implementing depolarizing errors for both intra-QPU and inter-QPU operations. 
Local gates experienced a base noise level $p_{\text{Local}}$, 
representing the probability that a gate operation applies an error to the qubit(s) it acts upon.
Specifically, under the depolarizing channel model, 
with probability $p$ a Pauli error 
is applied uniformly at random.
Non-local operations (communication primitives) experienced amplified noise $p_{\text{comm}} = \alpha \cdot p_{\text{Local}}$, 
where $\alpha \in [1.0, 1.1, 1.2]$ is the \textit{communication noise multiplier} representing the 
elevated communication error rates.
We varied Local noise levels from $0.001$ to $0.02$ and tested partition counts ranging from 2 to 6,
where \textit{partitions} refers to the number of sub-circuits (equivalently, quantum processing units) 
into which the original circuit is divided.
For a partition to be considered valid it must hold at least one qubit \footnote{
For small or low-depth circuits such as GHZ state preparation, extreme partitioning regimes (approaching one qubit per QPU) are algorithmically unnatural and are therefore best interpreted as stress tests of communication-induced noise and mitigation overhead.
}.
Each experiment configuration was executed with 200 measurement shots.
All non-local operations were implemented via quantum teleportation.
Due to the 31-qubit simulator limit, circuit-partition combinations exceeding this threshold
(primarily higher partition counts on larger circuits) could not be evaluated,
resulting in incomplete coverage for some algorithm-partition pairs.
However, the majority of the experimental design space was successfully sampled.

We evaluate mitigation strategies using three primary metrics:

\begin{itemize}
\item ZNE error: Expectation value error used to measure circuit fidelity, 
defined as the absolute difference between measured and ideal expectation values.
We report three quantities:
\begin{equation}
E_{\text{baseline}} = |\langle\hat{O}\rangle_{\text{measured}} - \langle\hat{O}\rangle_{\text{ideal}}|_{\text{no ZNE}}
\end{equation}
\begin{equation}
E_{\text{ZNE}} = |\langle\hat{O}\rangle_{\text{measured}} - \langle\hat{O}\rangle_{\text{ideal}}|_{\text{with ZNE}}
\end{equation}
\begin{equation}
\Delta E = E_{\text{baseline}} - E_{\text{ZNE}}
\end{equation}
where $E_{\text{baseline}}$ quantifies the error in the distributed circuit without mitigation, 
$E_{\text{ZNE}}$ measures the error after applying ZNE, 
and $\Delta E$ represents the absolute improvement.
Lower error values indicate better performance, 
while larger $\Delta E$ indicates more effective mitigation.

\item Error reduction: Fractional improvement in expectation value error when ZNE is applied:
\begin{equation}
\text{Error Reduction} = \frac{E_{\text{baseline}} - E_{\text{ZNE}}}{E_{\text{baseline}}}
\end{equation}
where $E_{\text{baseline}}$ is the expectation value error for the distributed circuit without mitigation, 
and $E_{\text{ZNE}}$ is the error after applying ZNE.
An Error Reduction of $-1.0$ implies that mitigation doubles the error, 
$0$ indicates no improvement, 
and $1.0$ corresponds to full correction.
In our analysis, we excluded $396$ outlier experiments arising from small baseline errors ($< 0.03$), 
for which the metric becomes numerically unstable.
In contrast to ZNE error, higher error reduction indicates improved performance.

\item Depth overhead: Ratio of the amplified circuit depth to the original circuit depth:
\begin{equation}
\text{Depth Overhead} = \frac{d_{\text{ZNE}}}{d_{\text{original}}}
\end{equation}
This metric captures the computational cost of ZNE and is critical for assessing whether 
error mitigation justifies the increased circuit depth, 
which reduces the coherence-limited execution window.
Since circuit distribution aims to partition large computations across resource-constrained devices, 
and ZNE inherently increases circuit size through noise scaling operations, 
monitoring depth overhead ensures that mitigation does not negate the size-reduction benefits of distribution.
\end{itemize}

\subsection{Circuit Partitioning}
To partition quantum circuits across multiple quantum processing units, 
we adopt a greedy community detection-based strategy that exploits the structure of qubit interactions.
The key idea is to group qubits that interact frequently into the same partition, 
thereby minimizing the number of gates that must be executed across partition boundaries.

We represent each circuit as an interaction graph, 
where vertices correspond to qubits and edges capture multi-qubit gate interactions.
Partitioning is then performed in three stages:
\begin{enumerate}
    \item Community identification:
    We first identify natural qubit communities using greedy modularity optimization applied to the interaction graph.
    This step clusters qubits with dense internal connectivity and sparse external interactions.
    
    \item Community adjustment:
    The resulting communities are iteratively merged or split to match a target partition count $k$.
    If the initial number of communities exceeds $k$, smaller communities are merged;
    if it is below $k$, larger communities are subdivided.
    
    \item Partition assignment:
    Finally, qubits are assigned to partitions according to the adjusted community structure.
\end{enumerate}

This hierarchical procedure balances two competing objectives:
reducing cross-partition gate cuts, which directly increase communication overhead,
and maintaining reasonably balanced partition sizes to avoid overloading individual devices.
Greedy partitioning has been shown to provide effective and computationally inexpensive circuit decompositions,
making it well suited for large-scale distributed quantum workloads \cite{g2021efficient}.

\subsection{Communication Primitives}

When a circuit is partitioned, gates that span partition boundaries cannot be executed directly 
and must be replaced with non-local communication primitives that enable inter-QPU quantum operations.
We use quantum teleportation (TP) as our communication primitive throughout this work, 
as it is the most widely studied and experimentally validated approach for DQC.

\begin{figure}
    \centering
    \includegraphics[width=1\linewidth]{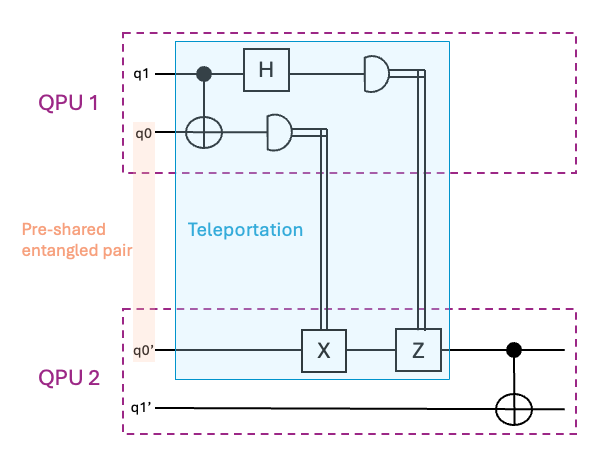}
    \caption{Teleportation (TP) implementation of a Local CNOT gate between two QPUs. Adapted from \cite{wu2022autocomm}. }
    \label{tpfig}
\end{figure}

Quantum teleportation~\cite{bennett1993teleporting} transfers an unknown quantum state from one processor to another 
using a pre-shared Bell pair (maximally entangled state) and two classical bits of communication.
To teleport the state of a qubit, the sender performs a Bell measurement between the qubit to be teleported 
and their half of the shared Bell pair, producing two classical measurement outcomes.
These outcomes are transmitted to the receiver, who applies corresponding Pauli corrections to their half of the Bell pair, 
recovering the original quantum state.
Figure~\ref{tpfig} illustrates a non-local TP based CNOT circuit where qubits q0 and q0' share the Bell pair and measurements of both q0 and q1 qubits of QPU1 are fowarded via classical channels to QPU2 to perform gates conditional on their result.

While teleportation preserves quantum information perfectly in the noiseless case,
it introduces several practical overheads in DQC, including:
\begin{itemize}
    \item Circuit depth: Each two-qubit gate crossing a partition boundary is replaced by a teleportation circuit 
    requiring additional gates (Bell state preparation, Bell measurement, and conditional Pauli corrections),
    increasing the total circuit depth.
    \item Qubit overhead: Each teleportation operation requires at least one ancilla qubit per QPU to store the Bell pair,
    meaning $k$ partitions require at least $k-1$ additional communication qubits beyond the original circuit qubits.
    \item Amplified noise: Entanglement generation and Bell measurements are typically higher-error operations 
    than Local gates, introducing elevated noise at partition boundaries 
    (which we model via our communication noise multiplier $\alpha$).
\end{itemize}

\subsection{ZNE Application Strategies}
In this work we compare two strategies for applying Zero Noise Extrapolation in distributed settings:

\begin{itemize}

    \item Global optimisation: ZNE is applied to the entire circuit before partitioning. 
The error-mitigated circuit is then partitioned, 
and cross-partition gates are replaced with TP primitives. 
This approach aims to preserve Global circuit structure during optimisation.

    \item Local optimisation: The circuit is first partitioned into sub-circuits, 
then ZNE is applied independently to each sub-circuit. 
This strategy aims to adapt mitigation to each partition's Local structure but may miss Global optimisations that span multiple partitions.

\end{itemize}

Both strategies use unitary folding for noise scaling with scale factors $\{1.0, 1.5, 2.0, 2.5, 3.0\}$ and linear extrapolation to the zero-noise limit. 

In neither case are the non-local operations mitigated.

Our experiments enable us to quantify the tradeoff between Global structural preservation and Local adaptive optimisation in the presence of partitioning overhead.

\section{Results}

\subsection{Scalability}

\begin{figure}
    \centering
    
    \begin{subfigure}{1\linewidth}
        \centering
        \includegraphics[width=0.9\linewidth]{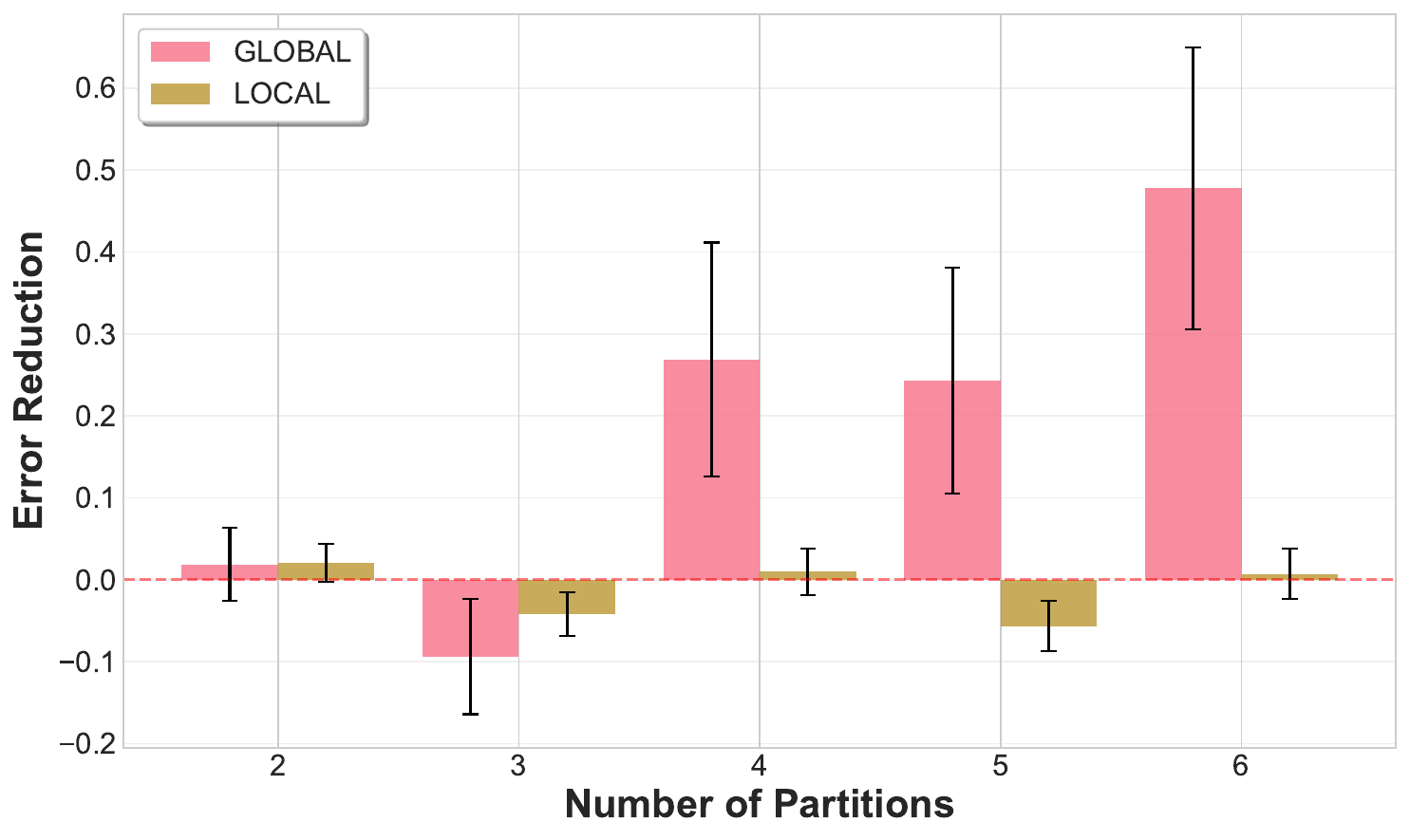}
        \caption{Error Reduction vs Partition}
        \label{fig:1a}
    \end{subfigure}
    
    \vspace{0.3cm}
    
    \begin{subfigure}{1\linewidth}
        \centering
        \includegraphics[width=0.9\linewidth]{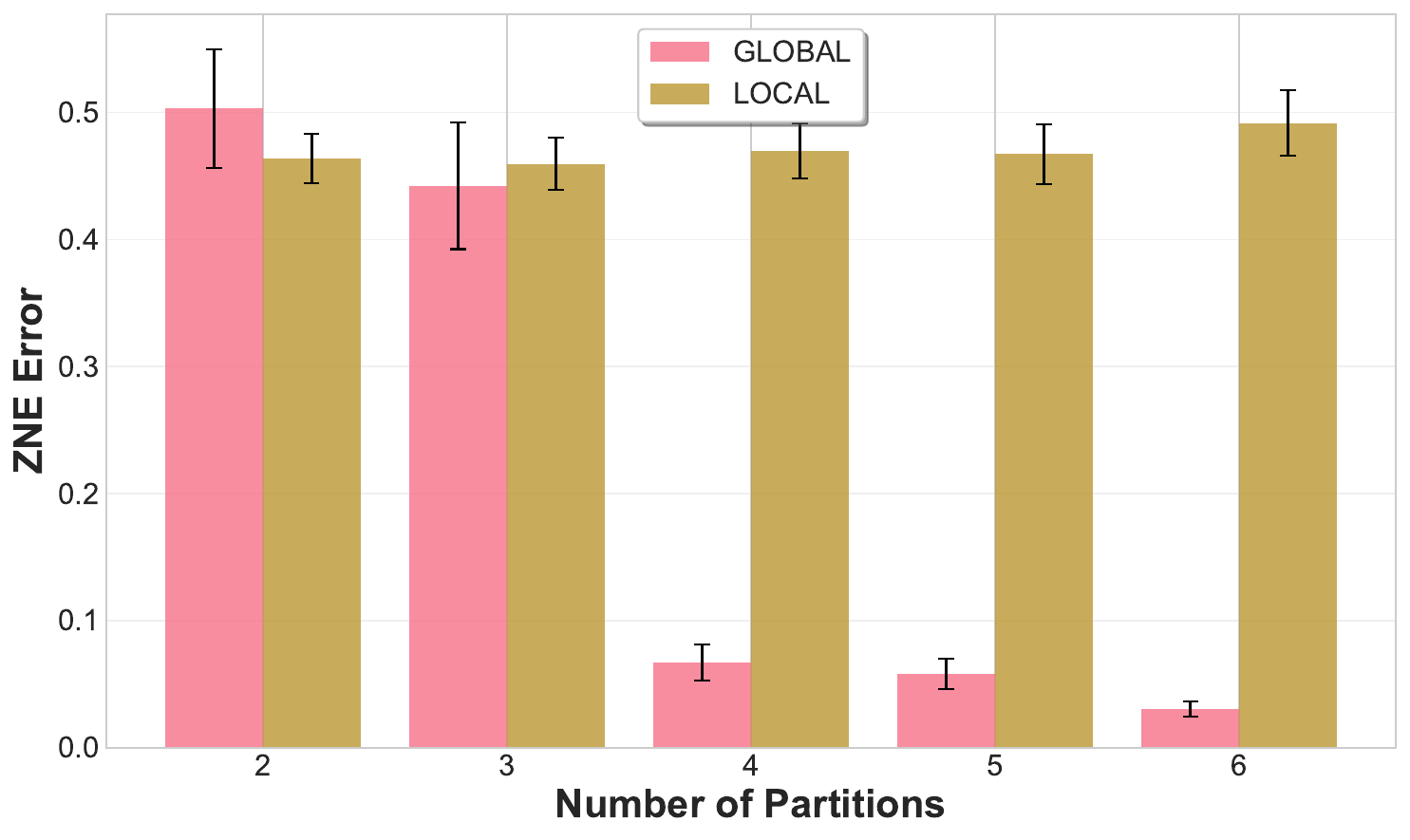}
        \caption{ZNE Error vs Partition}
        \label{fig:1b}
    \end{subfigure}
    
    \caption{Performance of ZNE strategies across partition counts, tested Local and communication error levels.}
    \label{fig:1}
\end{figure}

Global ZNE consistently achieves positive error reduction compared to the non-error-mitigated distributed baseline.
Performance improves considerably with increasing partition count (Figure~\ref{fig:1}).
Remarkably, this benefit persists despite additional noise from communication operations.
Global ZNE reaches an error reduction of roughly $48\%$ (at 6 partitions) when compared to a non-error mitigated noisy baseline. 

\begin{figure}
        \centering
        \includegraphics[width=0.9\linewidth]{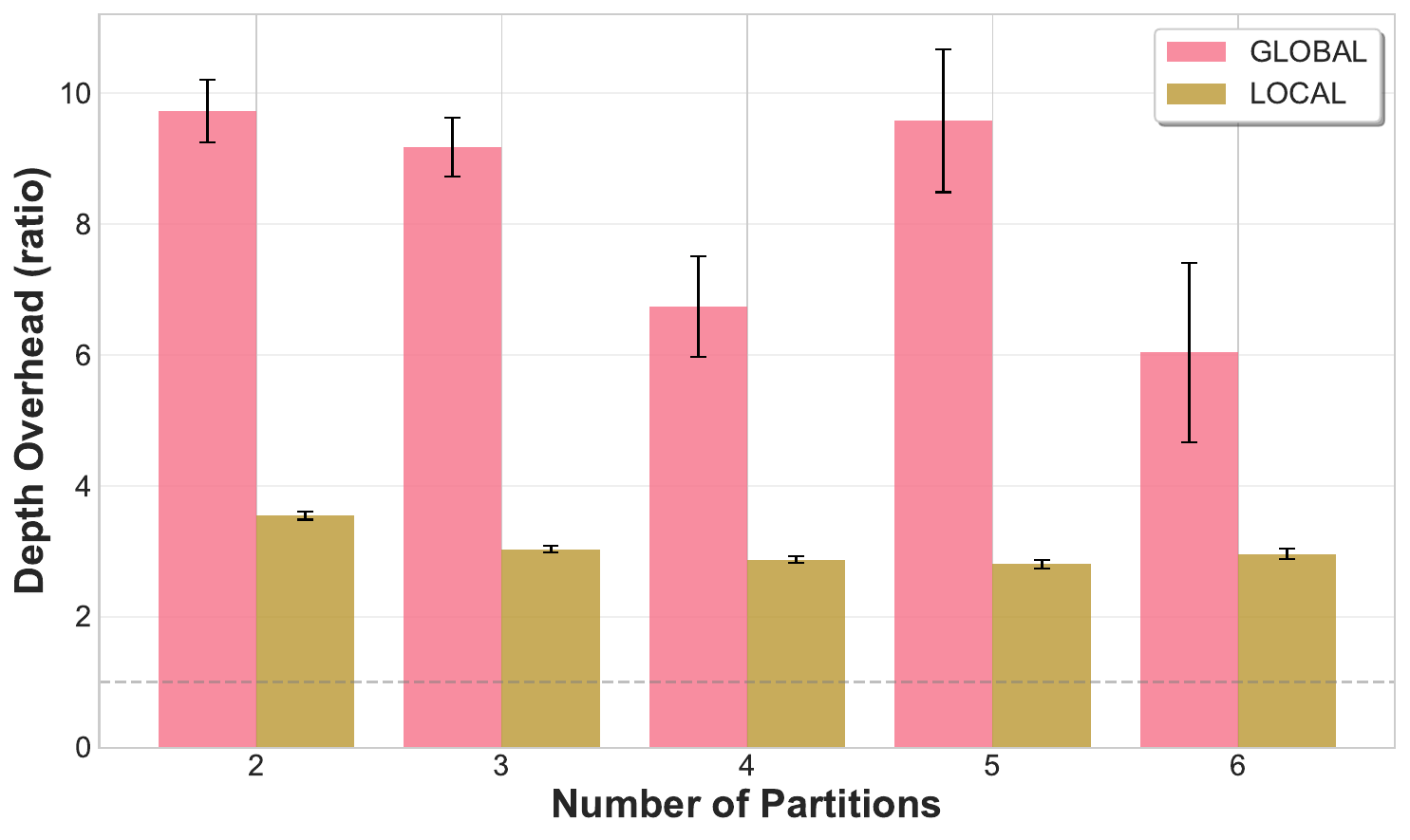}
        \caption{Circuit Depth Penalty across partition counts.}
        \label{fig:depth}
\end{figure}

This error reduction comes at a significant computational cost. 
Figure~\ref{fig:depth} shows that Global ZNE introduces substantial depth overhead, 
ranging from $6\times$ to almost $10\times$ the original circuit depth across different partition counts. 
Only at the maximum tested partition count of 6 does the per-device circuit depth of the Global encoding 
fall to that of the original monolithic circuit without error mitigation.

In contrast, Local ZNE shows inconsistent and often negative error reduction, 
with no clear scaling trend as partition counts increase.
Local ZNE maintains a more modest overhead of approximately $3\times$, 
which decreases slightly with increasing partitions. 
Only at 2 partitions does the per-QPU circuit depth 
exceed that of the non-mitigated monolithic circuit.

\subsection{Network Noise Resilience}

\begin{figure}
    \centering
    
    \begin{subfigure}{1\linewidth}
        \centering
        \includegraphics[width=0.9\linewidth]{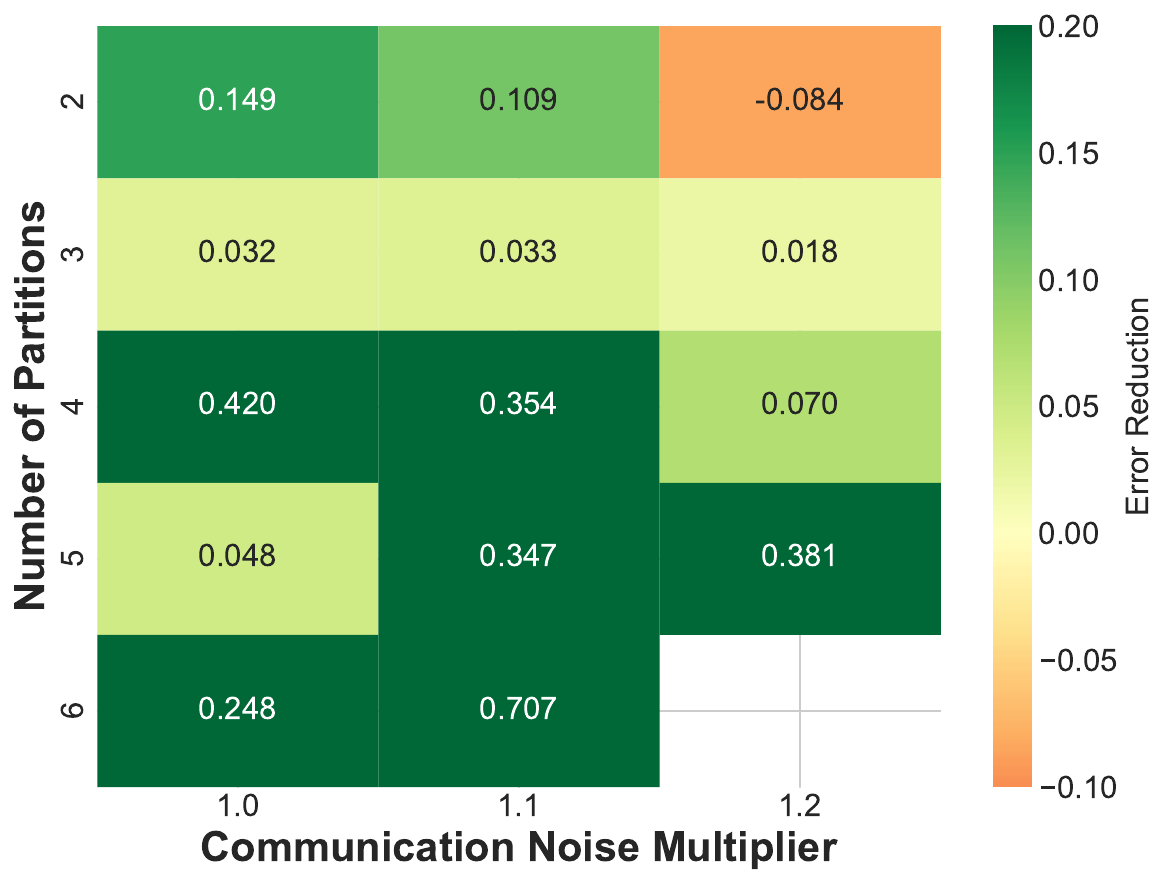}
        \caption{Global ZNE Strategy}
        \label{fig:2b}
    \end{subfigure}
    
    \vspace{0.3cm}
    
    \begin{subfigure}{1\linewidth}
        \centering
        \includegraphics[width=0.9\linewidth]{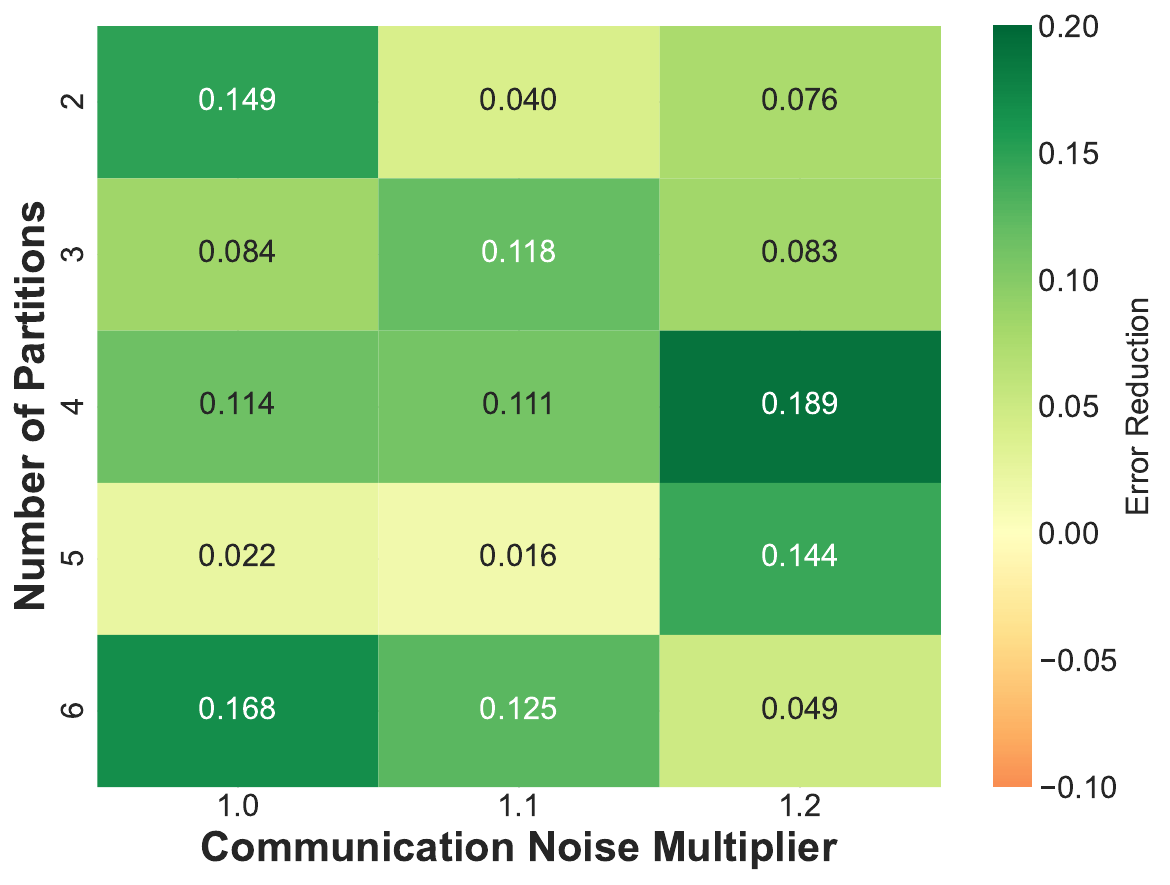}
        \caption{Local ZNE Strategy}
        \label{fig:2c}
    \end{subfigure}
    
    \caption{Error reduction heatmaps showing the interaction between partition count and communication noise multiplier. 
    Values represent mean error reduction (fractional improvement in expectation value error) after outlier removal using robust trimmed mean.
    Data is accounted from all tested local noise levels.}
    \label{fig:network_noise}
\end{figure}

The interaction between partition count and communication noise exhibits non-monotonic behaviour (Figure~\ref{fig:network_noise}). 
Global ZNE achieves substantial error reductions across most configurations, with the best performance of $71\%$ error reduction occurring at 6 partitions with moderate communication noise 
($\alpha =1.1$). 
At higher partition counts (4--6 partitions), Global ZNE maintains consistently strong performance, achieving error reductions exceeding $24\%$ even as communication noise increases.
Unexpectedly, performance degrades at lower partition counts (2--3 partitions), 
despite the reduced communication overhead that would be expected to be advantageous. 

Additionally, within high partition count configurations (5-6 partitions), elevated communication noise multipliers 
($\alpha =1.1$ and $1.2$) outperform the unrealistic baseline noise regime 
($\alpha =1.0$), particularly at partition counts above 4. 

In contrast, Local ZNE exhibits more uniform behaviour across partition counts, 
with performance remaining relatively stable across the parameter space. 
However, Local ZNE consistently underperforms Global ZNE, achieving only $1\%$--$19\%$ error reduction across all configurations tested.

\subsection{Local Noise Sensitivity}

\begin{figure}
    \centering
    
    \begin{subfigure}{1\linewidth}
        \centering
        \includegraphics[width=0.9\linewidth]{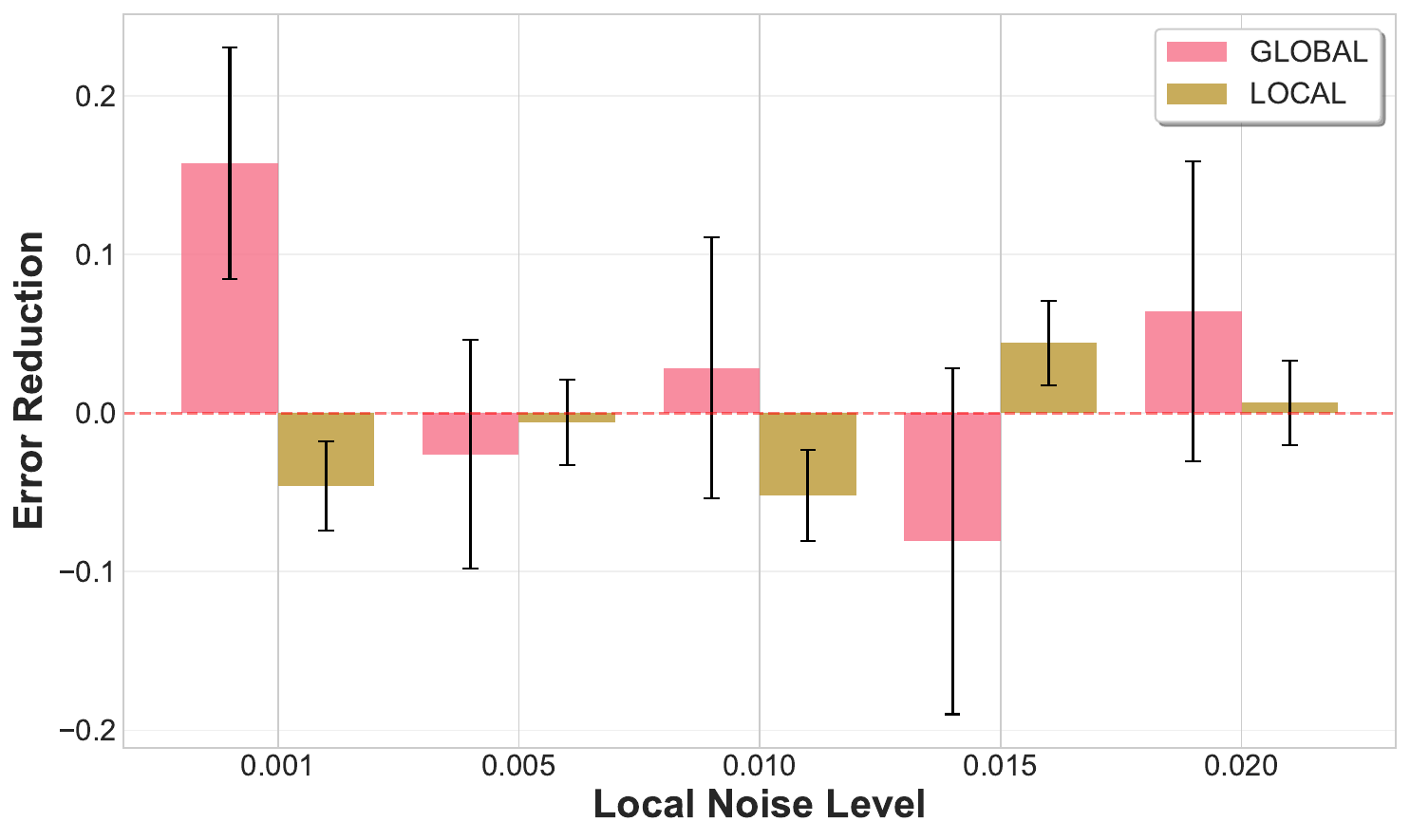}
        \caption{Performance vs Local Noise}
        \label{fig:3a}
    \end{subfigure}
    
    \vspace{0.3cm}
    
    \begin{subfigure}{1\linewidth}
        \centering
        \includegraphics[width=0.9\linewidth]{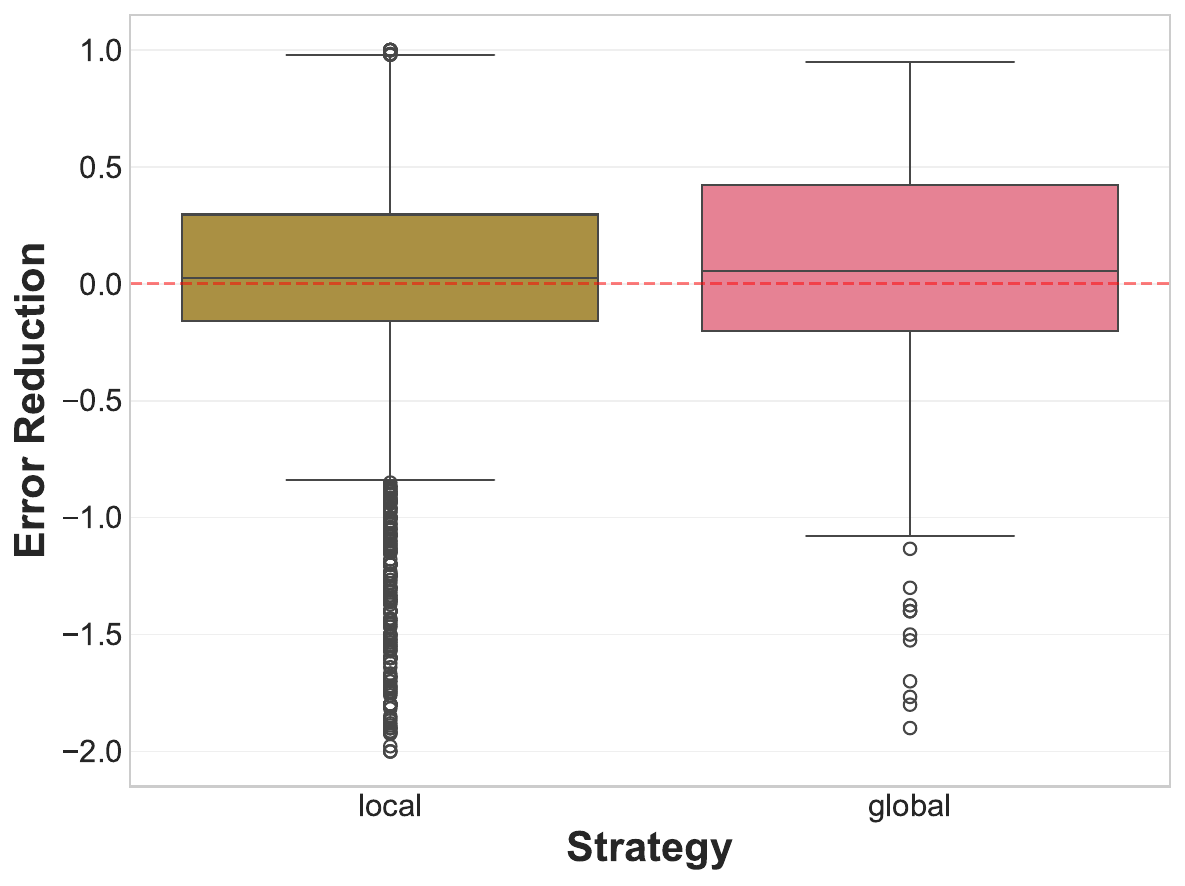}
        \caption{Strategy Performance Distribution}
        \label{fig:3b}
    \end{subfigure}
    
    \caption{Strategy comparison showing performance versus Local noise levels and overall performance distribution}
    \label{fig:strategy_comparison}
\end{figure}

Global ZNE achieves its highest error reduction ($\approx 0.16$) at the lowest tested noise level ($\epsilon = 0.001$), 
but exhibits highly variable performance as Local noise increases (Figure~\ref{fig:3a}). 
Contrary to expectations, error reduction does not decline monotonically with increasing noise. 
At intermediate noise levels ($\epsilon = 0.005$ and $0.015$), 
Global ZNE produces negative error reduction, indicating that ZNE amplification can amplify rather than mitigate errors under certain noise conditions.

Local ZNE, while achieving lower peak performance, 
shows substantially more stable behavior across noise levels. 
Error reduction remains near zero or negative throughout most of the tested range, 
with the notable exception of $\epsilon = 0.01$, where Local ZNE shows improved performance comparable to Global ZNE at that noise level.

The performance distributions (Figure~\ref{fig:3b}) quantify this stability trade-off. 
Global ZNE exhibits a higher median error reduction but with significantly wider variance, 
including numerous outliers in both positive and negative regions. In contrast, Local ZNE shows a tighter, 
distribution centered near zero, suggesting more reliable and modest, error mitigation across varying noise conditions.

\section{Discussion}

Our experimental results reveal complex interactions between error mitigation strategies, 
circuit partitioning, and noise characteristics in DQC. 
This section interprets the observed phenomena through four key lenses: 
scalability trends, computational overhead, unexpected noise behaviour, and practical implications.

\subsection{Global vs Local Optimisation}
The stark contrast between Global and Local ZNE performance suggests 
fundamentally different error modeling capabilities. 
We theorize that Global ZNE succeeds because noise amplification operates on the complete 
circuit topology, allowing the extrapolation model to capture error correlations that 
span partition boundaries. By preserving the original entanglement structure before 
partitioning, Global ZNE can model how errors propagate through the distributed system.

In contrast, Local ZNE treats each sub-circuit independently, effectively assuming the majority of 
errors are localized within partitions. This assumption breaks down when quantum communication 
introduces the expected $\alpha \cdot p_{\text{Local}}$ errors across QPU boundaries. 
These affect state fidelity throughout the distributed computation, not just within 
individual sub-circuits, making local encodings unable to capture cross-partition error correlations.

Interestingly, while Local ZNE achieves lower error reduction than Global ZNE, 
it shows remarkable stability across varying partition counts (Figure~\ref{fig:1}), 
maintaining consistent modest performance even as communication overhead increases (Figure~\ref{fig:2c}). 
This resilience suggests Local ZNE successfully mitigates errors within individual sub-circuits 
but lacks the circuit-wide perspective needed to model inter-QPU error propagation. 
The performance distributions (Figure~\ref{fig:3b}) further illustrate this trade-off: 
both strategies achieve similar median error reductions, but Global ZNE exhibits 
significantly wider variance, indicating higher risk alongside its higher potential reward.

An intriguing direction for future work would be to explore whether targeted error correction 
of network-induced noise, applied in conjunction with Local ZNE encodings, could approach 
or exceed the performance of Global ZNE. Such a hybrid approach might combine the computational 
efficiency and stability of Local mitigation with specialized correction techniques for 
communication primitives, potentially offering a more scalable alternative to Global optimisation 
as distributed systems grow in size.

\subsection{The Overhead-Resilience Trade-off}

ZNE's noise amplification introduces severe computational overhead through circuit depth inflation.
Global ZNE imposes a $6-10\times$ depth penalty while Local ZNE adds approximately $3\times$ 
overhead. 
At first glance, this appears paradoxical: 
circuit distribution aims to fit large computations within device capacity constraints, 
yet Global ZNE's inflation may eliminate this benefit by expanding circuit depth to unsustainable levels.

However, at the highest tested partition count, per-device circuit depth 
approaches the original monolithic circuit depth, despite the Global ZNE overhead,
potentially signalling to larger scale mitigation of this cost.

It is also worth noting that this trade-off will be present in quantum error correction encodings within distributed settings,
which may reach $10-100\times$ more physical qubits.

\subsection{Challenging the expected noise hierarchy}
Conventional intuition suggests that error mitigation performance should degrade as noise levels increase:
higher Local noise should reduce ZNE effectiveness, 
and larger communication noise multipliers should further amplify this degradation in distributed settings.
Our results defy this expectation.

Our data reveals highly non-intuitive behavior.
In several configurations, higher quantum network channel noise yields comparable or even superior error reduction relative to lower noise levels.
Similarly, higher partition counts sometimes correlate with improved performance, 
despite introducing additional communication primitives and, in principle, greater exposure to network noise.

Although the relationship between partition count and communication overhead is not strictly linear,
it remains strongly correlated:
increasing the number of partitions generally increases the number of inter-QPU gates,
thereby inflating circuit depth and communication noise exposure.
Under this model, the effectiveness of the mitigation would be expected to deteriorate.
Instead, we observe improved Global ZNE error reduction with increasing partition count,
suggesting that the pre-partitioning optimization captures structural properties of the circuit
that become increasingly beneficial at larger scales.
One possible explanation is that partitioning fragments the circuit into shorter subcircuits,
reducing coherent error accumulation and improving the numerical stability of ZNE extrapolation fits.
Moreover, the teleportation-based communication model considered here may alter the computation's noise structure
in ways that favor extrapolation, although it remains unclear to what extent this behavior will persist on real hardware.

\subsection{Limitations and Future Work}

Our study provides foundational insights into error mitigation for DQC, 
but several methodological constraints and unexplored research directions warrant acknowledgment.

\subsubsection{Simulator fidelity}
All experiments used the Qiskit Aer simulator with depolarizing noise models. 
While this enables controlled, reproducible experiments, 
real hardware exhibits correlated errors, crosstalk, gate-dependent error rates, 
and readout errors not captured by our simplified model. 
Hardware validation on actual distributed quantum platforms 
is essential to verify whether our findings translate to physical implementations.

\subsubsection{Algorithm coverage}
We evaluated three algorithm families from MQT Bench (Deutsch-Jozsa, GHZ, W state preparation).
Expanding our benchmarking suit to variational algorithms (such as QAOA, VQE), 
quantum chemistry implementations, and quantum machine learning algorithms 
would strengthen generalizability claims.

\subsubsection{Network realism}
Our model assumes perfect, instantaneous entanglement generation and ideal all-to-all connectivity.
Realistic networks have constrained topologies requiring multi-hop routing, 
rate-limited entanglement distribution with finite success probabilities, 
and purification overhead.
Incorporating these constraints would provide more realistic performance bounds.

\subsubsection{Mitigation technique scope}
We focused exclusively on Zero Noise Extrapolation with linear extrapolation. 
Other ZNE extrapolation variants (polynomial, exponential fitting, etc) 
and alternative mitigation techniques (probabilistic error cancellation, Clifford data regression) 
may interact differently with circuit partitioning.
Comparative evaluation across mitigation strategies remains critical future work.

\section*{Acknowledgment}
This research was supported by the EPSRC UK Quantum Technologies Programme under grant EP/T001062/1 and VeriQloud. I would like to thank the participants of WERQSHOP 2025 for their valuable feedback on early versions of this work. The code and data used in this work can be found at \href{https://github.com/grageragarces/ZNE-DQC}{\textcolor{linkblue}{github.com/grageragarces/ZNE-DQC}}.

\bibliographystyle{IEEEtran}
\bibliography{ref.bib}

\end{document}